\documentclass[10pt,conference,letterpaper]{IEEEtran}
\AtBeginDocument{%
  \providecommand\BibTeX{{%
    \normalfont B\kern-0.5em{\scshape i\kern-0.25em b}\kern-0.8em\TeX}}}
 
\usepackage{amsmath}
\usepackage{mathabx}
\usepackage{bbold}
\usepackage{array}
\usepackage{bm}
\usepackage{colortbl}
\usepackage{comment}
\usepackage{enumitem}
\usepackage{epsfig}
\usepackage[para]{footmisc}
\usepackage{graphicx}
\usepackage{wrapfig}
\usepackage[utf8]{inputenc}
\usepackage{multirow}
\usepackage{microtype}
\usepackage{subcaption}
\usepackage{tablefootnote}
\usepackage{tabularx}
\usepackage{url}
\usepackage{xcolor}
\usepackage{url}
\usepackage[numbers,sort&compress]{natbib}

\newcommand{\mas}{\glspl{mas}}

\usepackage[acronyms,nonumberlist,nopostdot,nomain,nogroupskip]{glossaries}

\begin{document}

\newacronym{rfid}{RFID}{Radio Frequency Identification}
\newacronym{rfp}{RFP}{radio fingerprinting}
\newacronym{sdr}{SDR}{software-defined radio}
\newacronym{mas}{MAS}{Mobile Autonomous System}
\newacronym{rl}{RL}{Reinforcement Learning}
\newacronym{5g}{5G}{fifth generation}
\newacronym{los}{LOS}{Line Of Sight}

\title{SeReMAS: Self-Resilient Mobile Autonomous Systems Through Predictive Edge Computing}
\author{\large Davide Callegaro$^\dagger$, Marco Levorato$^\dagger$ and Francesco Restuccia$^*$\\
\normalsize $^\dagger$Donald Bren School of Information and Computer Sciences, University of California at Irvine, United States\\
  $^*$Department of Electrical and Computer Engineering, Northeastern University, United States\\
\normalsize e-mail: \{dcallega,~levorato\}@uci.edu, frestuc@northeastern.edu
\vspace{-.3cm}

}


\maketitle

\begin{abstract}
Edge computing enables Mobile Autonomous Systems (MASs) to execute continuous streams of heavy-duty mission-critical processing tasks, such as real-time obstacle detection and navigation. However, in practical applications, erratic patterns in channel quality, network load, and edge server load can interrupt the task flow's execution, which necessarily leads to severe disruption of the system's key operations. Existing work has mostly tackled the problem with \emph{reactive} approaches, which cannot guarantee task-level reliability. Conversely, in this paper we focus on learning-based \textit{predictive} edge computing to achieve \textit{self-resilient} task offloading. By conducting a preliminary experimental evaluation, we show that there is no dominant feature that can predict the edge-MAS system reliability, which calls for an ensemble and selection of \emph{weaker} features. To tackle the complexity of the problem, we propose SeReMAS, a data-driven optimization framework. We first mathematically formulate a Redundant Task Offloading Problem (RTOP), where a MAS may connect to multiple edge servers for redundancy, and needs to select which server(s) to transmit its computing tasks in order to maximize the probability of task execution while minimizing channel and edge resource utilization. We then create a predictor based on Deep Reinforcement Learning (DRL), which produces the optimum task assignment based on application-, network- and telemetry-based features. We prototype SeReMAS on a testbed composed by a Tarot650 quadcopter drone, mounting a PixHawk flight controller, a Jetson Nano  board, and three 802.11n WiFi interfaces. We extensively evaluate SeReMAS by considering an application where one drone offloads high-resolution images for real-time analysis to three edge servers on the ground. Experimental results show that SeReMAS improves the task execution probability by $17\%$ with respect to existing reactive-based approaches. To allow full reproducibility of results, we share the dataset and code with the research community.\vspace{-0.2cm}
\end{abstract}


\glsresetall

\section{Introduction}

Mobile Autonomous Systems (MASs) such as self-driving cars and drones are disrupting the wireless, embedded and computing manufacturing industries, with unprecedented repercussions on agriculture, film making, surveillance and urban mobility, among others. According to a study by PwC, the current global market value for drones is estimated to be over \$127 billion \cite{mazur2016clarity}, while it is expected that North America's self-driving car market will expand at a CAGR of 50.8\% with a global revenue of \$49.79 billions by 2024 \cite{SelfDriving}. Thus, it is no wonder that \mas\  have captured the interest of academia and industry, now rushing to research and develop \mas-related devices and technologies across many different facets \cite{cao2019intelligent}. Characterized by the abundance of rich sensors (\textit{e.g.}, cameras, radars, and GPS), coupled with \gls{5g} wireless networking and advanced mobility, \mas\ are unique devices that can travel between destinations with little to no human control. To achieve this complex endeavor, \mas\ necessarily require the continuous, real-time execution of \textit{streams} of computation-expensive tasks. For example, self-driving cars have to continuously build detailed 3D maps of the surrounding areas, and use them to categorize different navigation features such as blockages, intersections, driveways, or fire hydrants. Moreover, autonomous drones are always at risk of sudden and significant drift due to adverse weather conditions, loss of power and/or GPS connectivity. Therefore, the seamless fusion of multimedia sensor data for real-time  path planning is quintessential for the drone's survival. \smallskip

\textbf{Motivation and Problem Setting.}~Offloading the stream of tasks generated by the \mas\  to edge servers can extend battery lifetime and reduce task round-trip time delay~\cite{bonomi2012fog}. However,  strong assumptions such as perennial stability of high-capacity communication links do not apply in the highly-dynamic context of \mas, where wireless links are bound to exhibit erratic behavior even in very simple scenarios. This key problem is further exacerbated in larger \mas\ and urban deployments, where parameters such as server and network load may induce more system instability. 

\begin{figure}[!h]
    \centering
    \includegraphics[trim = 10px 10px 10px 20px, width=\columnwidth]{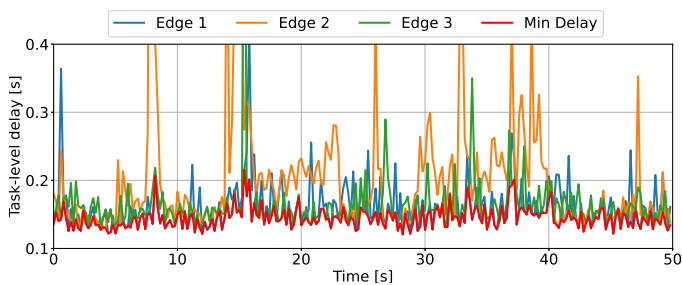}
    \caption{Example of task level delay from a flying drone to 3 edge servers, transmitted over WiFi 802.11n in a $50$s interval.\vspace{-0.3cm}}
    \label{fig:time_trace}
\end{figure}

In this paper, we tackle the challenging problem of providing \emph{task-level} performance guarantees to a stream of computing tasks generated by an airborne MAS. Specifically, we impose a bound on the maximum time between data acquisition and the completion of the corresponding analysis task. We remark how a task-level perspective is necessary in this class of systems, where temporally local degradation of task delay can severely harm control loops in \mas.
Figure \ref{fig:time_trace} shows the temporal pattern of the end-to-end application delay (the different curves are different edge servers) obtained through our experimental drone testbed described in Section~\ref{sec:exp_setting}. We can observe that the delay exhibits significantly time-varying patterns, with a standard deviation 0.14 and a peak-to-peak difference reaching
0.43, which is 241\% of the average value of 0.178s.
We note that the experimental setting is in \gls{los}, and that more convoluted propagation environments would just aggravate this problem. A bound on the average delay would not guarantee that the task-level delay will be below a certain threshold for \emph{each} of the tasks belonging to the task stream, which is key to guarantee correct functionality of stream-oriented edge-based \mas.

Our vision is simple: \textit{the seamless usage of edge resources by \mas\ necessarily requires techniques able to mitigate the impairments and erratic temporal patterns induced by the surrounding communication and computing ecosystems and the physics of the system itself}. Existing work -- discussed in detail in Section \ref{sec:rw} -- has tackled the issue of \mas\ reliability in a piecemeal and often highly abstract fashion, by focusing on static optimization of either mobile device's trajectory \cite{zhou2018computation,cheng2018uav,zhang2018stochastic,hu2018joint} or communication resources \cite{yang2019energy,zhang2020joint,bertizzolo2020swarmcontrol}. In Sec.~\ref{sec:exp_setting}, we show that edge selection methodologies based on channel quality would fail, and we conclude that new task offloading strategies are needed to stabilize task completion delay in \mas.

To address this challenging problem, we developed SeReMAS -- Self-Resilient Mobile Autonomous Systems -- a framework whose core is a dynamic task replication mechanism, where individual tasks are replicated and sent over multiple channel/edge server resources. The key intuition is that the task delay experienced by the MAS will be the minimum delay of each replica. Thus, the larger the number of channel/edge couples, the greater the probability that one task will satisfy the delay requirement, which however also implies increased resource usage. The objective of SeReMAS is to minimize resource usage under the constraint that the probability that the task-level delay bound will be met.\smallskip

\textbf{Our Approach.}~To drive our design, we implemented a testbed composed by an airborne MAS and multiple ground servers (Section \ref{sec:prel_exp}). Specifically,  we extracted a rich dataset from the system (Section \ref{sec:exp_setting}), whose analysis demonstrates a lack of variables strongly correlated with the delay (Section \ref{sec:prelim}). We show that the received signal strength indicator (RSSI), one of the key variables used to control connectivity and offloading, has limited influence on the delay. The dataset illustrates how in real-world MAS systems the delay pattern is the result of a wide variety of complex cross-variable interactions at various temporal scales. Importantly, influential variables are outside the network layers, and include \emph{physical} variables such as orientation, acceleration and tilt.

\begin{figure}[!h]
    \centering
    \includegraphics[width=.95\columnwidth]{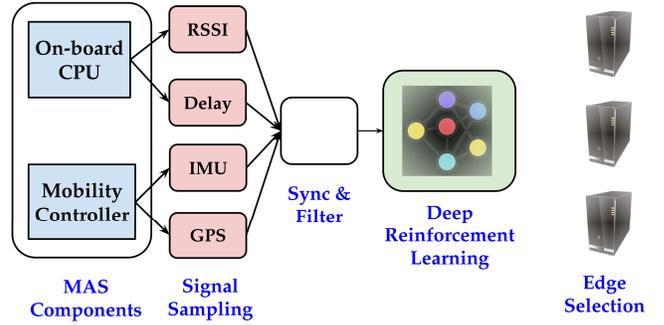}
    \caption{Our Architecture for Task Offloading in MASs.}
    \label{fig:self_aware_architecture}
    \vspace{-0.4cm}
\end{figure}

Based on this considerations, SeReMAS embeds a \emph{predictive} core based on Deep Reinforcement Learning (DRL) to determine a compact set of computing pipelines dynamically assigned task-by-task based on the perceived state of the system. Fig.~\ref{fig:self_aware_architecture} depicts the high-level schematics of SeReMAS. The key intuition is that the selected set of channel/computing resources will influence future decision making, which DRL is able to capture. Some of the features -- e.g., application and most network-related features -- become available only if a resource is used. For instance, if a channel/edge server pipeline is not selected for a task, then the corresponding delay is not observed, which motivates the adoption of a DRL-based approach. By including future rewards in action selection and taking as input unprocessed features such as RSSI, end-to-end delay, inertial measurement unit (IMU) and global positioning system (GPS) coordinates, \textit{the DRL algorithm will implicitly embed the impact of current computing pipelines selection on the efficacy of future decisions, as well as real-world phenomena that can be hardly modeled through explicit mathematical terms}.\vspace{-0.2cm}

\subsection*{\textbf{Novel Contributions}}

$\bullet$
We design SeReMAS, a framework for the dynamic  control of task offloading in \mas\ with extreme temporal variations (Section \ref{sec:middleware}). SeReMAS is based on a preliminary experimental analysis (Section \ref{sec:prelim}), which indicates that there is no dominant feature, including obvious features such as channel quality, and that prediction necessitates an ensemble of \emph{weaker} features. We first mathematically formulate (Section \ref{sec:problem_formulation}) a Redundant Task Offloading Problem (RTOP). Then, we create predictors that can help managing the resource usage/performance trade-off. Specifically, we propose a myopic predictor as baseline (Section \ref{sec:myopic}) and a DRL-based approach, which operates on a set of features from application, network and device-level components (Section \ref{sec:dq_formulation}). \textit{To the best of our knowledge, SeReMAS is the first framework addressing the problem of redundant task offloading in MAS with a data-driven approach which efficacy is verified in a real-world testbed and with replicable dataset-based experiments.}

\smallskip

$\bullet$ We prototype SeReMAS on a drone-based experimental testbed (Section \ref{sec:prototype}). The platform embeds a module for the real-time analysis of features, including the flight controller, tied to internal data routing control. As part of our prototype, we design a strategy to make the state representation compact (Section \ref{sec:fss}), and thus lower the complexity of the DRL agent, using an iterative feature selection procedure. We consider a real-time image analysis application through state-of-the-art edge-assisted object detection algorithms where a drone periodically acquires from onboard sensors data whose analysis is offloaded to edge servers on the ground (Section \ref{sec:exp_setting}). We let the drone perform task offloading through multiple WiFi interfaces, and collect a total of $140$ minutes of flying. \emph{The dataset and the code produced as part of this paper can be found at \cite{seremasgithub}.}

\smallskip

$\bullet$ Through experiments, we show how different subsets of features appear dominant at different time-scales (Section \ref{sec:exp_results_myopic}). We also show in Section \ref{sec:exp_results_redundant} how the DRL approach improves by 17\% the task execution probability with respect to a reactive approach \cite{paper_rssi}, thanks to the ability to manage state uncertainty in the action selection problem, measured in terms of probability of meeting a delay requirement per amount of resource used, with respect to a myopic controller based on a one-shot selection of the next set of edge servers to be used.


\section{Preliminary Experiments}\label{sec:prel_exp}

In our setting, a MAS is connected to $N$ edge servers ${\rm es}_1,{\rm es}_2,\ldots,{\rm es}_N$ through separate wireless channels. The device generates a sequence of tasks $t_1, t_2, t_3, \ldots$ with fixed inter-arrival time equal to $T$ seconds. A task is described as a chunk of data to be processed with a predetermined analysis algorithm to produce an output. We assume that tasks are homogeneous, meaning that the amount of data associated with any task and the analysis algorithm are fixed. Let us define $\delta_n(t_i)$ as the capture-to-output delay of task $t_i$ executed as edge server ${\rm es}_n$, defined as the time from the generation of the task to the availability of its output at the edge server. The delay $\delta_n(t_i)$ is the composition of two delays: the transmission delay $\delta^{\rm comm}_n(t_i)$ and the computing delay $\delta^{\rm comp}_n(t_i)$. In real-world settings, both components are highly stochastic, and depend on a number of latent variable, parameters as well as states of protocols at various layers of the stack.

\subsection{Preliminary Analysis}
\label{sec:prelim}

\begin{figure}[!h]
    \centering
    \begin{subfigure}[b]{0.49\columnwidth}
        \centering
        \includegraphics[trim= 10px 0 10px 0, width=\textwidth]{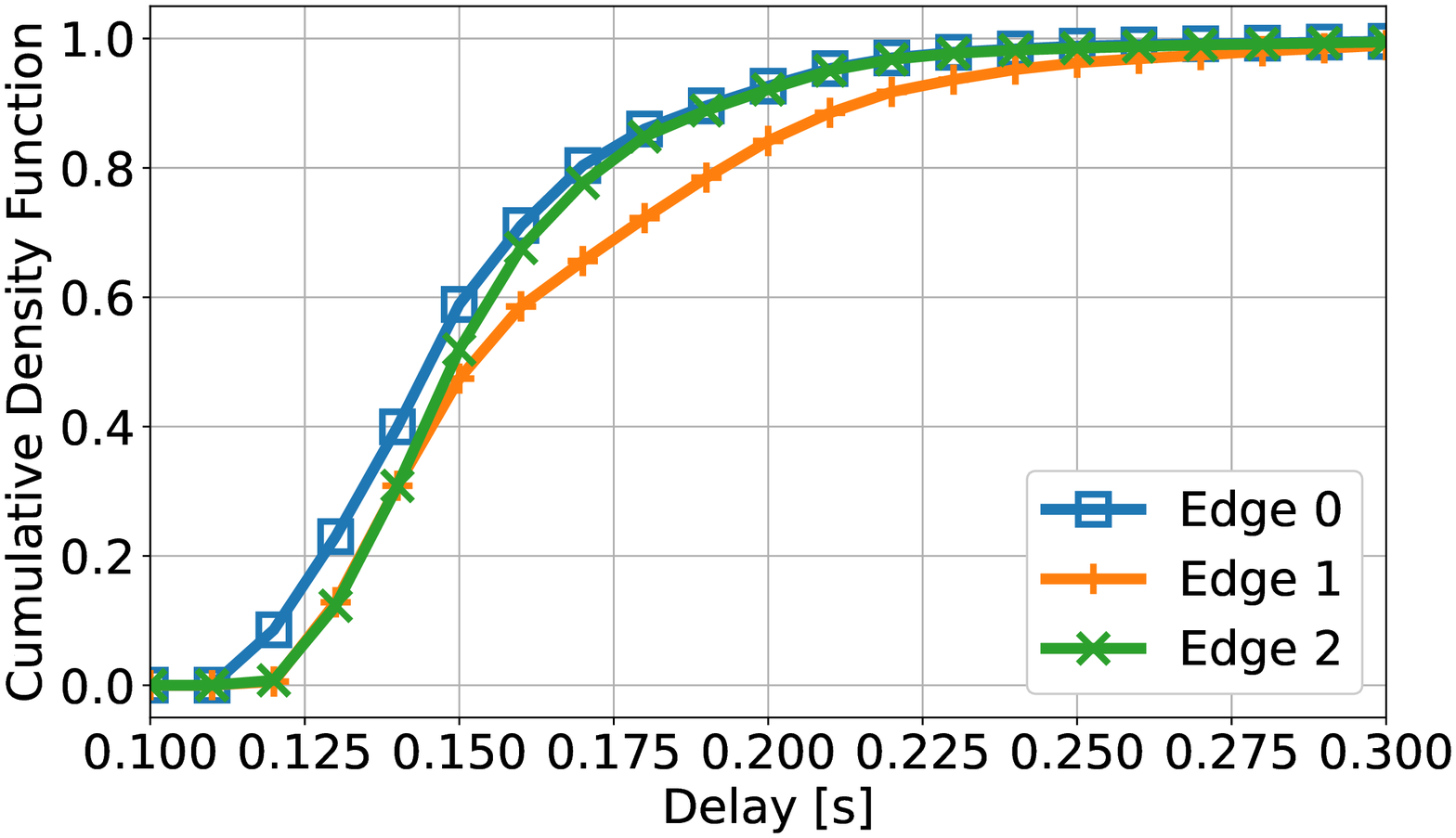}
        \caption{}
    \end{subfigure}
    \begin{subfigure}[b]{0.49\columnwidth}
        \centering
        \includegraphics[trim= 10px 0 10px 0, width=\textwidth]{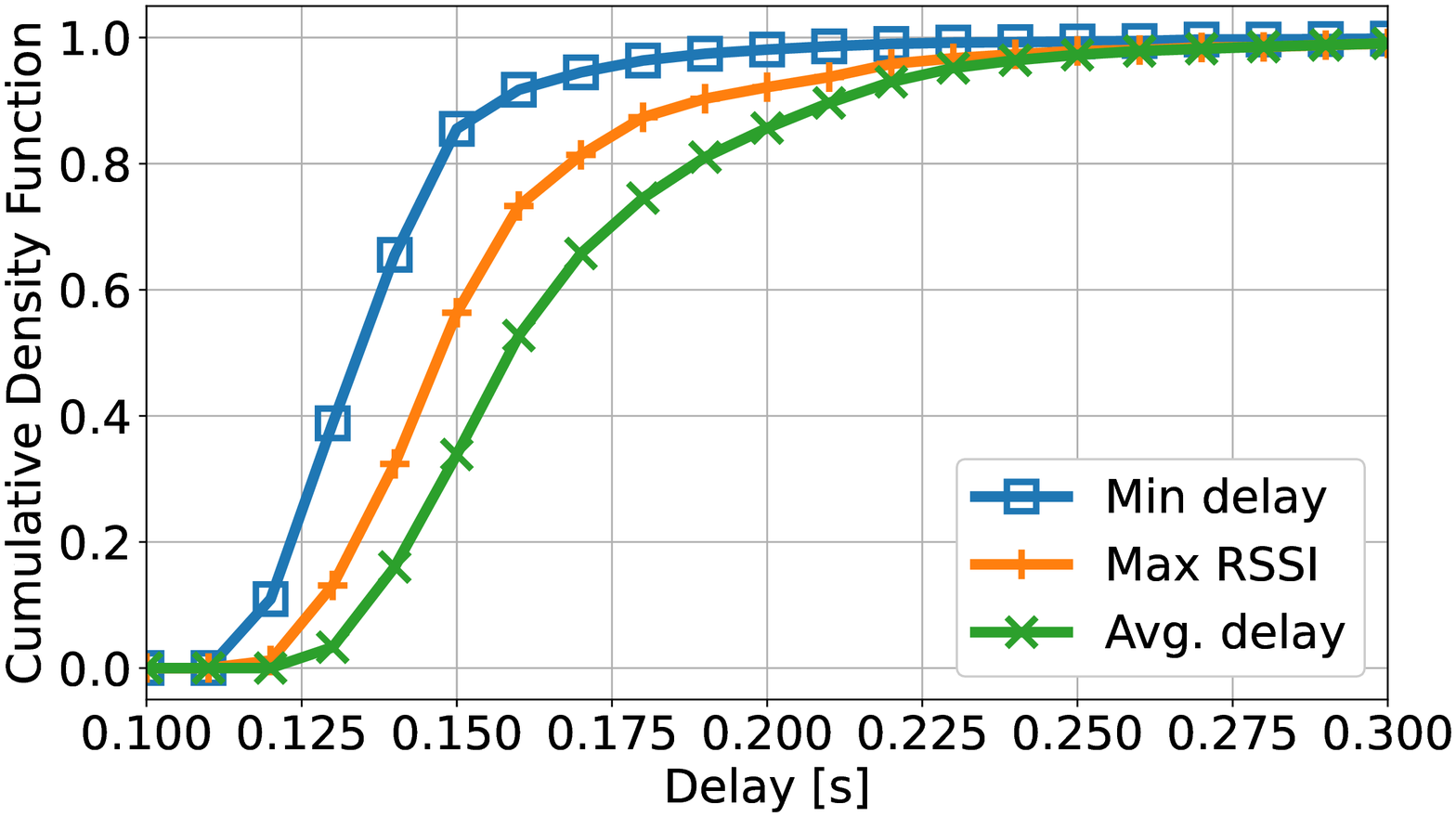}
        \caption{}
    \end{subfigure}
    \caption{ Cumulative density function of delay (a) for each edge server and (b) selecting the minimum delay, or the one with maximum RSSI, or the average of the available delays.\vspace{-0.3cm}}
    \label{fig:delay_distr}
\end{figure}

We motivate our study by analyzing the data obtained from real-world experiments. We consider an experimental setting, described in detail in Section \ref{sec:exp_setting}, where a drone is offloading image processing tasks to three edge servers.  Fig.~\ref{fig:time_trace} shows a section of the temporal pattern of the task-level delay $\delta_t$ at the three edge servers. We observe that the delay signals alternate low-delay ($150-175$ms) sections with spikes and higher delay sections. While some mild correlation between the delay signals is present, the minimum of the three signals provides the needed stability to the delay. Fig.~\ref{fig:delay_distr}.a shows the Cumulative Density Function (CDF) of the task-level delay $\delta_t$ for the three edge servers in our experiments. Note that in our scenario the task execution delay $\delta^{\rm comp}$ is nearly deterministic. We remark that all the edge servers are within coverage, and that all the links are in Line of Sight (LoS). Most delays are in the range $120$ms to $250$ms, with about $40$\% of the delays below $135$-$145$ms.

Fig.~\ref{fig:delay_distr}.b shows the distribution of the minimum delay $\delta_{\rm min}$ with respect to the cdf of the average delay and the delay associated with the edge server with the maximum channel quality index (RSSI). We observe that there is a noticeable difference between the minimum delay and the delay offered by the edge server with the best channel quality. Therefore, even a perfect SNR-based handover would fail to provide optimal performance in this context. This effect is the result of the convoluted interdependencies between protocol variables at the various layers and the physical and hardware properties of the system at multiple time scales.

\begin{figure}[!h]
    \centering
    \begin{subfigure}[b]{0.49\columnwidth}
        \centering
        \includegraphics[trim= 10px 0 10px 0, width=\textwidth]{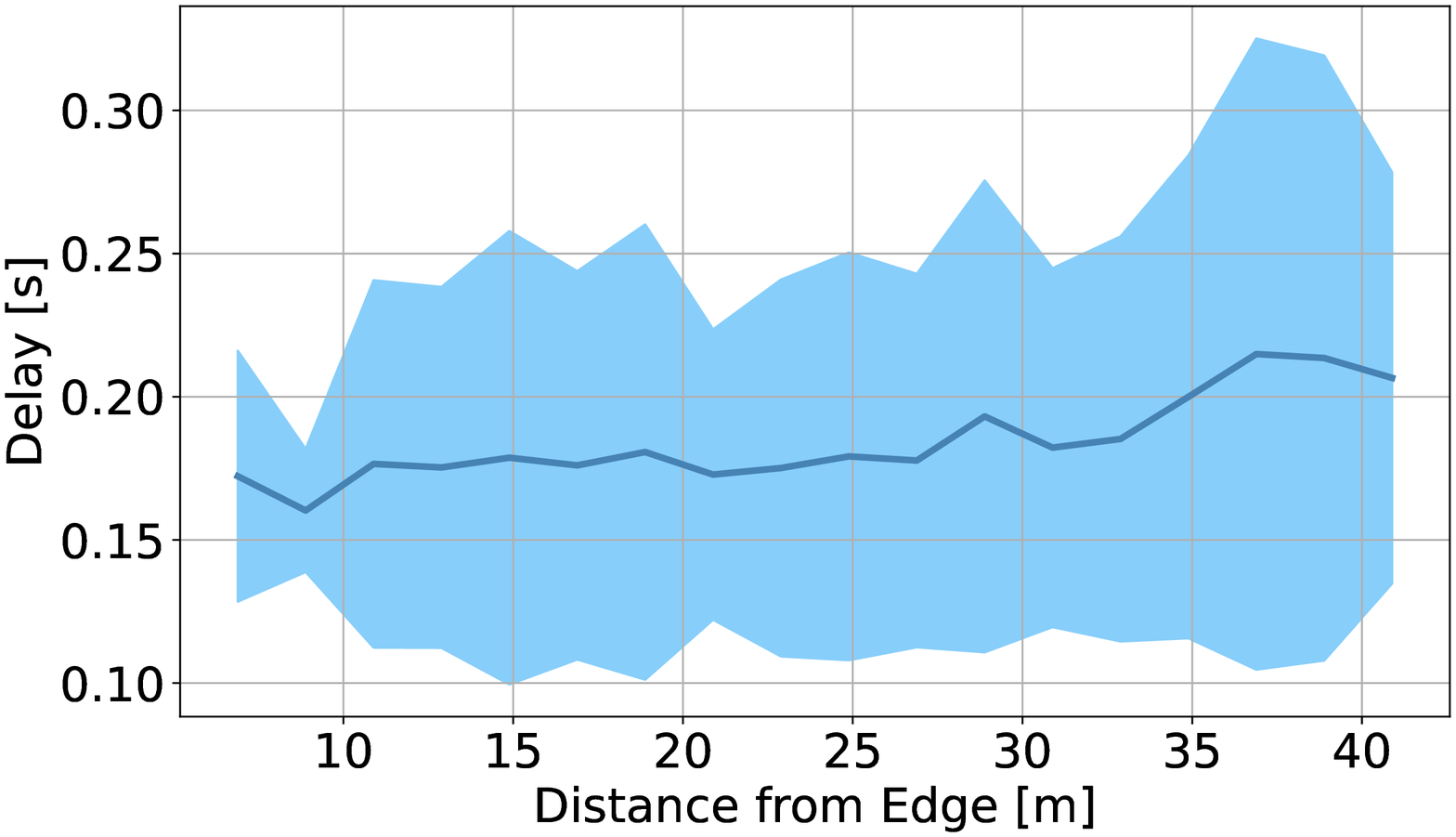}
        \caption{}
    \end{subfigure}
    \begin{subfigure}[b]{0.49\columnwidth}
        \centering
        \includegraphics[trim= 10px 0 10px 0, width=\textwidth]{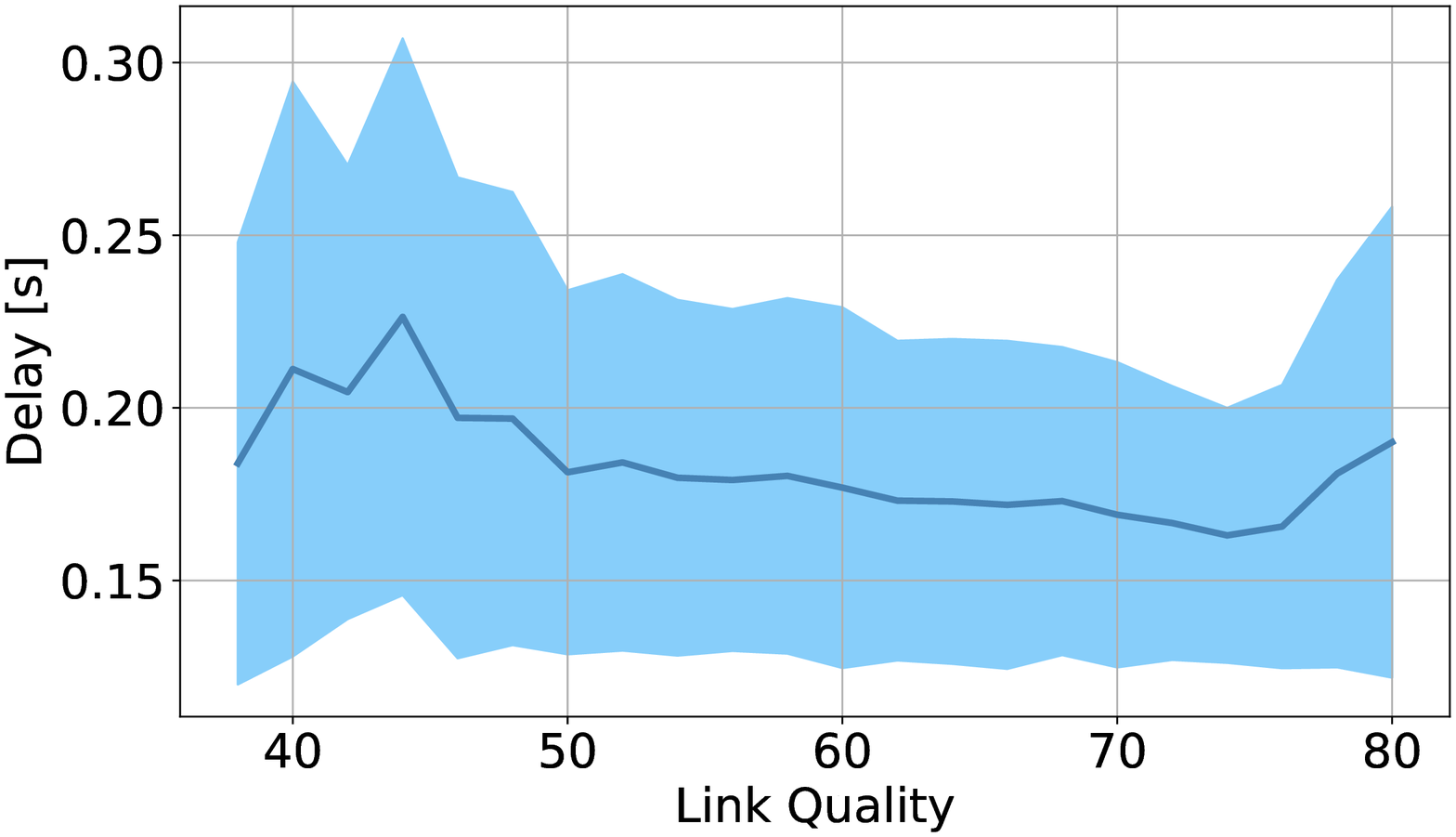}
        \caption{}
    \end{subfigure}
    \caption{ Distribution of task level delay as a function of distance from each of the edges and the RSSI. 
    \vspace{-0.3cm}}
    \label{fig:delay_distance_rssi}
\end{figure}

We remark this important aspect by plotting in Fig.~\ref{fig:delay_distance_rssi} the (delay, RSSI) and (delay, distance) mean and one standard deviation of the delay as a function of the two other variables. We can see the lack of a strong correlation between the delay and both RSSI and distance and emphasize again how experimental results unveil effects and interactions that are rarely captured in simulations and models.

\section{The SeReMAS Framework}\label{sec:middleware}

The results illustrated in the previous section emphasize the need for new techniques boosting the reliability of edge offloading for extreme real-time applications. In this section, we present SeReMAS, a data-driven framework addressing the reliability of task offloading in MAS. We first present an overview of the main system blocks and functionalities in Section \ref{sec:walkthrough}. Then, we formalize the learning-based redundant task offloading control problem in Section \ref{sec:problem_formulation}.

\subsection{SeReMAS: A Walkthrough}\label{sec:walkthrough}

\begin{figure}[!h]
         \centering
         \includegraphics[width=\columnwidth]{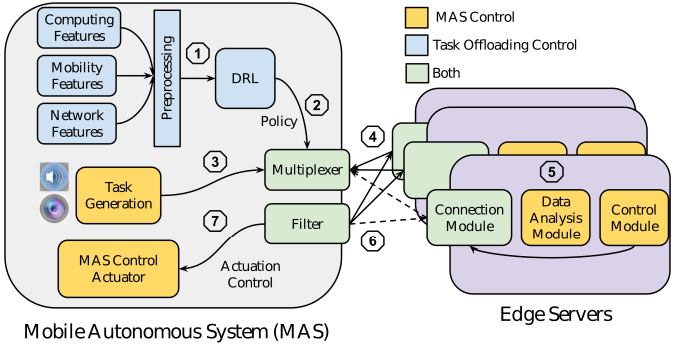}
         \caption{SeReMAS system architecture: two different control cycles intersect at the communication modules, where the DRL agent's policy is applied by means of task replication.}
        \label{fig:architecture_new}
\end{figure}

SeReMAS~\cite{seremasgithub} enables the data-driven control of task offloading from the MAS to the edge servers. The architecture of SeReMAS is depicted in Fig.~\ref{fig:architecture_new}, where we show the modules performing mobility control of the MAS (yellow) and control of task offloading (blue), and the modules -- multiplexing and filter -- handling the communication between the section of the platform at the MAS to the section at the various edge servers. 

We now provide a walk-through of the main operations performed by SeReMAS, following the steps indicated  in Fig. \ref{fig:architecture_new}.
First, the framework takes computing features (e.g., CPU, GPU, and RAM utilization), mobility features (e.g., accelerometer, gyroscope, GPS coordinates, etc), and network features (e.g., TCP state, RSSI) and applies pre-processing (step 1) to construct the input to a DRL model (see Section \ref{sec:problem_formulation} for details). The extracted features and the composition of the state space are described in Section~\ref{sec:fss}. Then, the DRL state is given as input to the DRL algorithm, which outputs $\phi$, the set of edge servers to be used as task executors (step 2).

Tasks are generated (step 3) according to the current MAS needs (e.g., multimedia classification), and handled by module called \textit{multiplexer} (step 4) which handles task replication across multiple edge servers. Specifically, the multiplexer is responsible for replicating and forwarding the tasks to the edge servers, and is directly controlled by the output ${\bf \phi}$ of the DRL algorithm. The tasks are sent to the edges specified by ${\bf \phi}$, which are then executed (step 5). The knowledge produced by the task execution can be used to drive control decisions on the MAS. 
For example, in our prototype we use the task result to control the mobility of the MAS, as explained in Section \ref{sec:prototype}. 
The related control messages generated by the edge server(s) are sent back to the MAS, and processed by the filter module (step 6), which eliminates replicated messages when more than one edge server is selected to avoid the re-execution of flight commands.
Finally, the control messages are fed to the control actuator (step 7), which takes care of implementing the control action, if needed (e.g., flight control).

\subsection{Redundant Task Offloading Problem (RTOP)}
\label{sec:problem_formulation}

As part of the SeReMAS framework, we investigate the problem of redundant task offloading to replicate tasks and send them over multiple channel-edge server pipelines for increased reliability, which we call RTOP. This problem will drive our DRL design.
We define the capture-to-output delay as the minimum of the delays associated with the task replicas:
 \begin{equation}
    \delta_{t_i}(\phi_i) = \min\{\delta_n(t_i):n{\in}\phi_i \},
\end{equation}
where $\phi_i{\subseteq}\{1,\ldots,N\}$ is the subset of edge servers to which a replica of task $t_i$ is sent. 
Then, we define a controller whose objective is to determine the sequence of edge servers $\boldsymbol{\phi}^*{=}[\phi_{t_1}^*,\phi_{t_2}^*,\ldots]$ solving the following optimization problem:
\begin{align}
    \arg\min_{\boldsymbol{\phi}} &~~\mathit{E}_i\left[|\phi_i|\right]  \\
    {\rm s.t.}&~~ \mathit{E}_{i}\left[  \mathit{I}\left( \delta_{\rm min}(t_i)>\delta^* \right) | \phi_i \right]<\Delta,
\end{align}
where $\mathit{I}(\cdot)$ is the indicator function and expectation is computed over the task sequence. This formulation is different than imposing a constraint on the average delay, i.e.,  $\mathit{E}_{i}\left[  \delta_{\rm min}(t_i) | \phi_i \right]<\delta^*$. The latter formulation would allow a possibly large number of delays above $\delta^*$, while our formulation is equivalent to
 \begin{align}
    \arg\min_{\boldsymbol{\phi}} &~~ \mathit{E}\left[|\phi_{t}|\right]   \\
                      {\rm s.t.} &~~P\left( \delta_{\rm min}(t)>\delta^* | \phi_{t}\right) <\Delta.
\end{align}
Thus, we impose a constraint on the probability that the task completion time is above a threshold $\delta^*$ while striving to minimize resource usage.

Intuitively, the larger the number of edge servers selected, the larger the probability that the minimum of the delays is below the threshold. However, the inevitable limitations on channel access and maximum edge server load leads to a \emph{task-level selection} problem, where the number and members of the chosen set is informed by the uncertainty regarding future delays and their expected values.  In real-world settings, the resolution of the RTOP defined above necessitates the consideration of complex inter-variable and temporal interdependencies. For this reason, we resort to data-driven solutions methodologies decomposing the problem into sequences of local problems.

\subsection{Myopic-based Baseline for RTOP}\label{sec:myopic}

First, we formulate a myopic predictive solution to address the RTOP. We introduce the notion of state of the system $s_i=\{s_{i,n}\}_{n=1,\ldots,N}$, where $s_{i,n}$ is the feature matrix
\begin{equation}
\label{eq:feature_matrix}
    s_{i,n} = \begin{bmatrix} 
\psi_{1,i-L+1,n} & \ldots & \psi_{1,i-1,n} & \psi_{1,i,n} \\ 
\psi_{2,i-L+1,n} & \ldots & \psi_{2,i-1,n} & \psi_{2,i,n} \\ 
          \vdots & \vdots & \ldots & \vdots\\
\psi_{F,i-L+1,n} & ... & \psi_{F,i-1,n} & \psi_{F,i,n}
\end{bmatrix},
\end{equation}
of $F{\times}L$ features, and $\psi_{f,j,n}$ is $f-th$ feature referring to task $j$ and computing pipeline $n$. We describe the specific features and dataset in the Section~\ref{sec:fss}. We train a probabilistic predictor as the function $p_{i+1,n} = \sigma(s_{i,n})$, where
\begin{equation}
    p_{i+1,n} = P\left( \delta_{n}(t_{i+1})>\delta^* \right).
\end{equation}

We  find the set $\phi_{i+1}$ with minimum cardinality such that
\begin{equation}
    P\left( \delta_{t_{i+1}}(\phi_{i+1})>\delta^*\right) <\Delta,
\end{equation}
where the left-hand term is computed as
\begin{equation}
    1{-}\prod_{n{\in}\phi_{i+1}}(1{-}p_{i+1,n}).
\end{equation}
When more than one set with the same cardinality satisfies the constraint, then the one with the smaller probability is chosen. We note that stronger predictors $\sigma(\cdot)$ may lead to a reduced resource usage, as they would lead to reduced uncertainty in the class of the next delay (above and below threshold), and thus would allow the controller to bet on fewer remote computing pipelines. For example, let us assume that at least one of the pipelines has a next delay below threshold: an accurate and confident predictor returning probability $1$ would allow the selection of only one edge server.



We extend the predictor to larger temporal windows to evaluate the predictive power of features blocks. We define
\begin{equation}
    p^{W,y}_{i+1,n} = \sigma(s_{i,n})
\end{equation}
where
\begin{equation}
    p^{W,y}_{i+1,n}=P\left(\sum_{\ell=0}^{W-1}\mathit{I}(\delta_{n}(t_{i+\ell})>\delta^*){\geq}y\right),
\end{equation}
that is $p^{W,y}_{i+1,n}$ is the probability that at least $y$ tasks will be completed with delay larger than $\delta^*$ in a window of $W$ future tasks.
We build a binary classifier from $\sigma(\cdot)$ by setting
\begin{equation}
   p^{W,y}_{i+1,n} \underset{C_0}{\overset{C_1}{\gtrless}} \frac{1}{2}
\end{equation}

\subsection{Deep Q-Learning Approach for RTOP}
\label{sec:dq_formulation}

\begin{figure}[h!]
    \centering 
    \includegraphics[width=0.9\columnwidth]{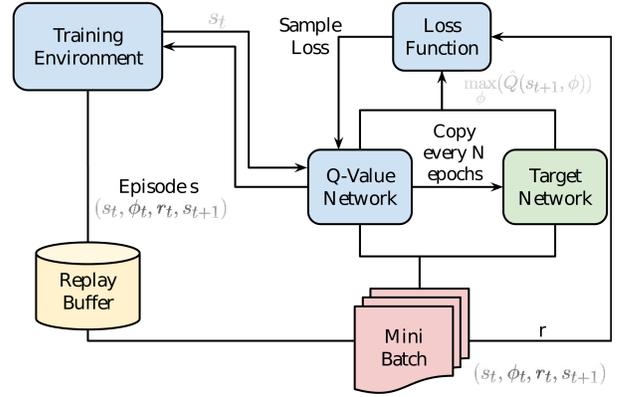}
    \caption{ Training architecture using Double Deep Q-Learning.}
    \label{fig:DDQN}
    \vspace{-4mm}
\end{figure}

The formulation above produces suboptimal control sequences.  Thus,  we adopt a Deep Q-Learning formulation to resolve the optimization problem. This formulation implicitly accounts for the impact of current decisions on the distribution of future states (and thus on the accuracy of control). In this case, the predictive function is defined to return the Q-values based on the state, that is,

\begin{equation}
    Q(s_{i+1},\phi_{i+1}) = \sigma_{DRL}(s_i),
\end{equation}
where
\begin{align}
    Q(s_i,\phi_i) &= \mathit{E}_{s_{i+1}|s_i,\phi_i}\left[\mathit{E}_{r_{i+1}|s_{i+1},\phi_i,s_i}\left[ r_{i+1} | s_{i+1},\phi_i,s_i| \right]\right]\nonumber \\ & + \gamma \max_{\phi^{\prime}} \mathit{E}_{s_{i+1}|s_i,\phi_i}\left[ Q(s_{i+1},\phi^{\prime}) \right].
    \label{eq:QV}
\end{align}
The cost variable $c_i$ includes weighted penalties for the delay being above threshold and the cardinality of the selected set, that is
\begin{equation}
\label{eq:cost}
    c_i = \lambda~ c_i^{\rm delay} + (1{-}\lambda)~c_i^{set},
\end{equation}
with
\begin{equation}
    c_i^{\rm delay}{=}\mathit{I}(\delta_{\min}(t_i)>\delta^*)
    \mathit{S}(\alpha^{\rm delay} \delta_{\min}  - \kappa^{\rm delay})
\end{equation}
and
\begin{equation}
c_i^{\rm set}= \alpha^{\rm set}|\phi_i| - \kappa^{\rm set},
\end{equation}
where $\alpha^{\rm delay}$, $\alpha^{\rm set}$, $\kappa^{\rm delay}$ and $\kappa^{\rm set}$ are normalization and offset parameters. $\mathit{S}(x){=}1/(1+e^{-x})$ is the sigmoid function, here used to generate a smooth delay cost function which is $0$ until $\delta^{*}$ and then progressively penalizing higher delay without overpenalizing tasks with poor channel conditions. Figure \ref{fig:DDQN} shows the training procedure for our DRL-based approach.

We remark that the recursive formulation of the Q-values embeds the distribution of future states and costs given the current policy.
The Q-values guide the selection of the actions according to the rule:
\begin{equation}
\phi_{i+1} = 
\begin{cases} 
    \text{argmin}_{\phi_i} Q(s_i, \phi_i) & \text{with prob. } 1-\epsilon_t \\
    \mathcal{U}(\mathcal{P}(\phi) \setminus \emptyset) & \text{with prob. } \epsilon_t
  \end{cases}
\end{equation}
where the best action (that is, subset of servers) is selected as the one maximizing the future reward with probability $1-\epsilon$, and selected uniformly at random with probability $\epsilon$.
This is commonly known as a \textit{$\epsilon$-greedy} strategy, it is often used in practical problems to balance exploration/exploitation in DRL problems.

\section{SeReMAS Prototype}\label{sec:prototype}

We first describe the platform experimental components in Section \ref{sec:platform}, and then describe our feature selection process in Section \ref{sec:fss}.
Finally, we explain how we implemented the SeReMAS predictors for the RTOP, both myopic and DRL, in Section \ref{sec:dql_impl}.

\begin{figure}[!h]
    \centering
    \begin{subfigure}[b]{0.49\columnwidth}
        \centering
        \includegraphics[width=\textwidth]{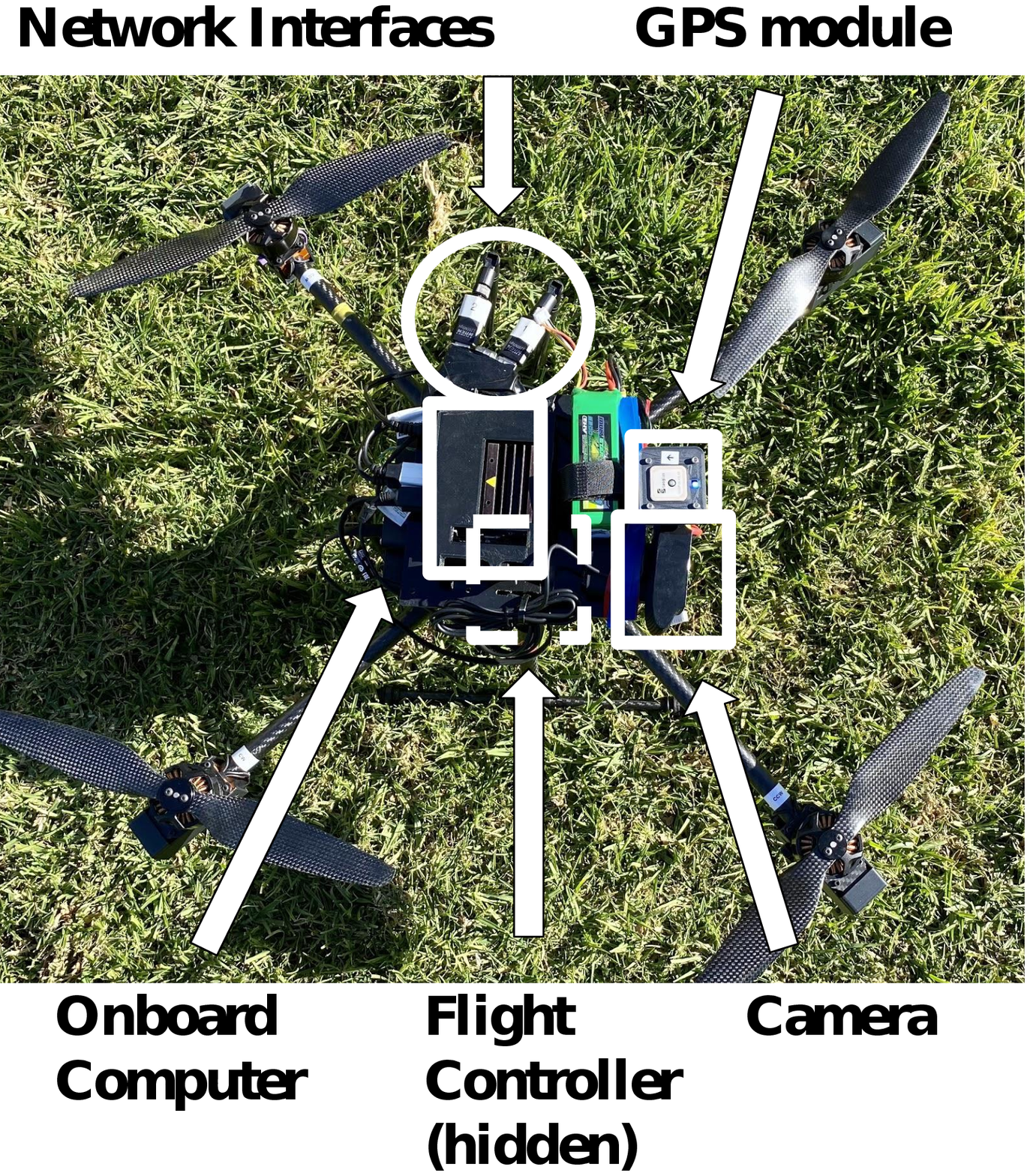}
        \caption{}
    \end{subfigure}
    \begin{subfigure}[b]{0.49\columnwidth}
        \centering
        \includegraphics[width=\textwidth]{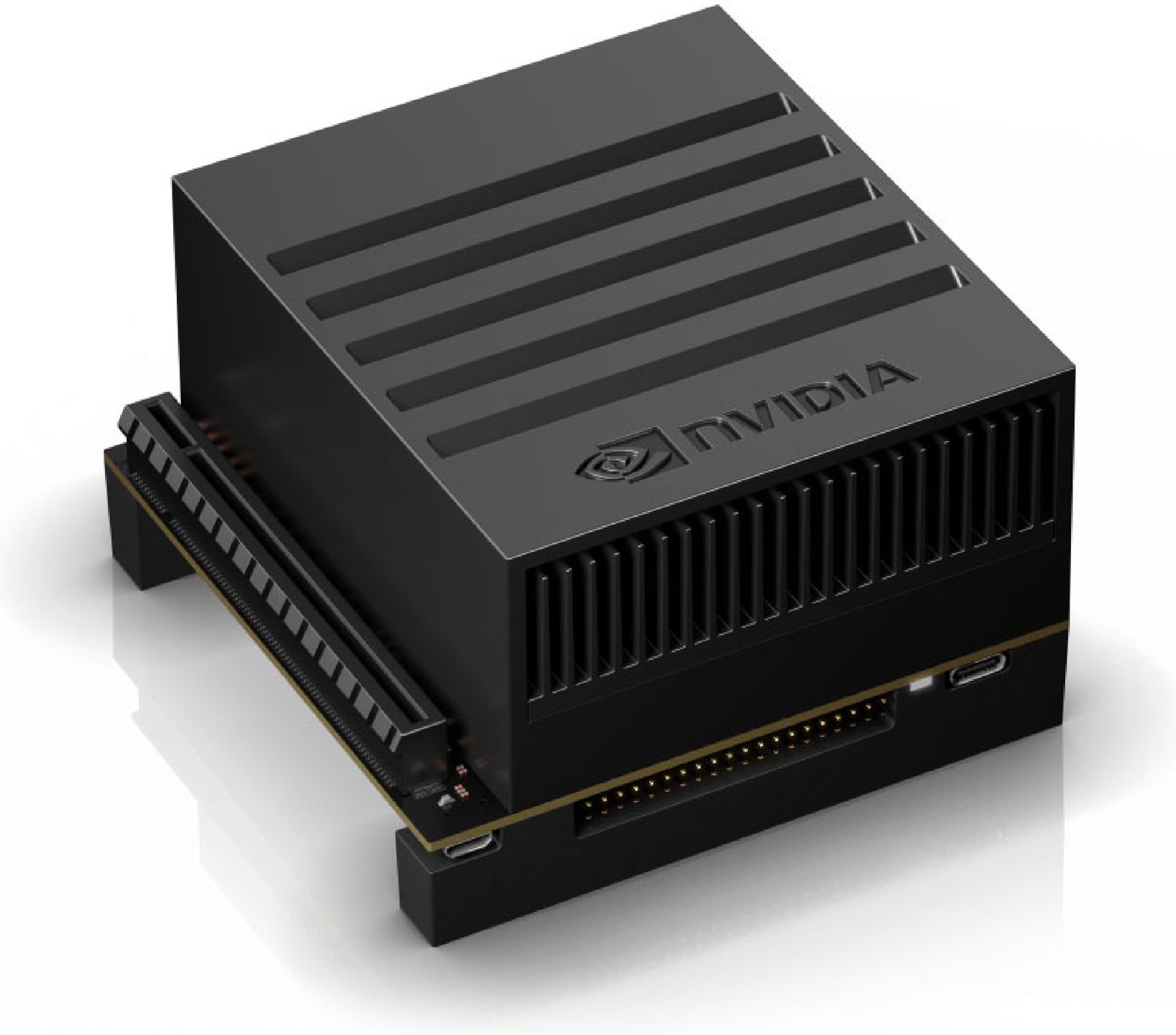}
        \caption{}
    \end{subfigure}
    \caption{(a) Drone prototype; (b) NVidia Jetson Xavier acting as edge server. \vspace{-0.2cm}}
    \label{fig:boards}
\end{figure}

\subsection{Platform Components}\label{sec:platform}

Figure \ref{fig:boards} shows our experimental setup. Specifically, we use a Tarot650 quadcopter mounting a PixHawk flight controller. 
We connect Telem2 port on the PixHawk to a serial interface on a NVidia Jetson Nano board with $4$GB of RAM. We use three NVidia Jetson Xavier development boards, operating in performance mode with 8 core ARM 64-bit processor, 32GB of main memory, 512-core Volta GPU. We use three IEEE 802.11n WiFi cards to interconnect the drone to the edge servers. These boards act as access points on different channels in the 2.4GHz WiFi spectrum.


\subsection{DRL State Space and Feature Selection}\label{sec:fss}

We discuss how we create the input state for the DRL algorithm. We consider features at the application, network stack and device level as follows:

\begin{itemize}
\item {\bf Application and Onboard Computer:} We track relevant application variables such as past capture-to-control delays, number of samples in the intermediate buffers, and selected actions. These will include real-time statistics relative to power consumption and resource allocation of CPU, GPU, and RAM.

\item {\bf Telemetry and Position: }We use MAVLink \cite{8743355} protocol messages to register a listener to the flight controller. The onboard computer receives monitoring statistics from the Inertial Measurement Unit (IMU), Global Positioning System (GPS) and the power consumption of the vehicle.
We include the edge servers' position, by including the distance from the drone using polar coordinates (Distance, Azimuth, Elevation) centered in the reference edge server. Distance is computed using the Harvesine formula.
Moreover, we add the relative heading by computing the orientation of the drone with respect to the position of the edge server.
Furthermore, we consider the $L2$-norm of multi-dimensional vectors (such as accelerometer and gyroscope data) and compute speed with respect to absolute reference frame and edge servers. All the features are synchronized at $5$Hz.  

\item {\bf Network:} We select relevant parameters such as TCP window and retransmissions, RSSI, and modulation/coding scheme (MCS) of the IEEE 802.11n protocol. We do so separately for each network interface available, so to isolate features relative to each edge server.

\end{itemize}

The details of the features are available in~\cite{seremasgithub}. \smallskip

\textbf{Feature Selection.}~We use feature importance methods such as Logistic Regression, Support Vector Machines and Random Forest as implemented in \cite{pedregosa2011scikit} and selected Logistic Regression with L$1$ regularizer due to the bias that Random Trees have towards features with high support's cardinality and the hybrid nature of the features, which include continuous and categorical variables.
We then used a recursive algorithm, where at each iteration we train a predictor and discard the least influential features.
We reduce the initial pool of $360$ features to $73$, maximizing accuracy on the validation set. Table I shows the normalized feature relevance  predicting the number of high-delay tasks in a $1~s$ window.

\begin{table}[!h]
\begin{center}
\begin{tabular}{| l | c |}
\hline
Feature & Normalized Correlation \\
\hline
 Round Trip Time average & 1\\  
 Transmission timeout & -0.83 \\
 Packets Received & -0.80 \\
 Channel Level & -0.48 \\
 Inclination (magnitude) & -0.17 \\
 Position w.r.t Edge & 0.16 \\
 Altitude & 0.16 \\
 Last Sent & -0.15 \\
 Heading & 0.13 \\
 Speed & 0.08 \\
 Congestion Window & 0.08 \\
 \hline
\end{tabular}
\caption{Normalized feature relevance to a linear model predicting the number of high-delay tasks in a $1~s$ window.\vspace{-0.6cm}}
\label{table:1}
\end{center}
\end{table}

Interestingly, while all available past delays are selected in the prediction (with $L=3$ in Eq. \ref{eq:feature_matrix}), acceleration and inclination features are selected with a lag of $0.6$s indicating a longer range dependency with the delay. Other relevant features include gyroscope and the increment of TCP fast retransmissions, failures, RSSI, and retries. The complete trend within the window is selected for these features.
The selection shows how both vehicle and network parameters are relevant to characterize the state of the system and its future behavior, but their influence is expressed at different time scales.

\subsection{Myopic Predictor and DRL Implementation}
\label{sec:dql_impl}

We provide the details of the myopic and DRL controllers.

\vspace{1mm}
\noindent
{\bf Myopic Predictor - }
To implement the predictor $p^{W,y}_{i+1,n} = \sigma(s_{i,n})$ we train a series of dense DNNs (with two hidden layers at $[150, 50]$ nodes) using the Adam optimizer and trained for $100$ epochs, with softmax output), which returns the probability that the next delay will belong to the predicted class.

\vspace{1mm}
\noindent
{\bf Deep Q-Learning Agent - }
Naive implementations of Deep Q-Learning use one DNN function. However as demonstrated in~\cite{ddqn}, this approach may cause instability during training if the Q-values presents sudden changes. Due to the erratic behavior of the system we consider, we then take a Double Deep Q-Learning (DDQL) approach to build our DRL agent. In DDQL, two separate Deep Neural Networks (DNN) are used. Referring to Eq.~(\ref{eq:QV}), one network is trained to approximate $Q(\cdot)=Q(s_i,\phi_i)$, and the other one to approximate the future Q-value term in the expectation, that is $\hat{Q}(\cdot)=Q(s_{i+1},\phi^{\prime})$

Fig.~\ref{fig:DDQN} illustrates the DDQN architecture and the training procedure. 
We use a fully connected DNN, with [$200, 100, 50$] hidden nodes, ReLu activation, and Huber Loss.
During training, we apply backpropagation to $Q(\cdot)$ over the epochs $e=1,...,N$. We periodically copy DNN's parameters so that $\hat{Q} \leftarrow Q$, as a mean to reduce noise in during training.
Note that we still choose the best action to learn on $\phi$ using the most updated $Q(\cdot)$, and in fact the decoupling between action selection and q-value function evaluation further stabilizes learning. We use a replay buffer during training, where the experiences in the form of $(s_i, \phi_i, r_i, s_{i+1})$ are stored and sampled randomly to avoid forgetting, which may occur if only the most recent experiences are used~\cite{dnn_forgetting}.

\section{Experimental Results}
\label{sec:exp_result}

We first present the experimental setting in Section \ref{sec:exp_setting}, then the prediction performance in Section  \ref{sec:exp_results_myopic}, and the task offloading results in Section \ref{sec:exp_results_redundant}.

\subsection{Experimental setting}
\label{sec:exp_setting}

\begin{figure}[h!]
    \centering
    \includegraphics[width=.95\columnwidth]{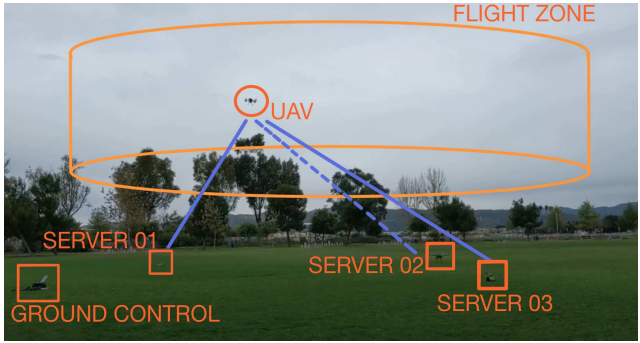}
    \caption{Schematic representation of the system setting: three ground edge servers, connected to the drone. Not all connections are continuously actively used (unused is dashed).}
    \label{fig:system}
    \vspace{-4mm}
\end{figure}

We consider a testbed illustrated in Fig.~\ref{fig:system}, which is composed of an airborne drone and $N{=}3$ ground edge servers in \gls{los}.  We consider an object tracking application where the MAS uses a camera to follow a predefined object at a certain distance. Specifically, the MAS captures images that are analyzed to extract the bounding box of the closest object of a certain class (\emph{e.g.}, a person). 

The controller steers the vehicle in the appropriate direction to (\emph{i}) center the bounding box in the field of vision and (\emph{ii}) obtain a bounding box of a predefined size by controlling the distance with respect to the object.  In our testbed, the drone generates a regular stream of images to be analyzed using object detection. Specifically, the drone emits $15$ images of size $19.5~kB$ per second. 
SSD-MobileNet-v2 model is used to analyze the images. 
In our measurements, the NVidia Jetson Xavier board takes $10~ms$ to execute the algorithm. Note that the onboard NVIDIA Jetson Nano takes $87~ms$ to complete the execution, however, power expenditure increases from $1.6~W$ to $4.2~W$ when the GPU is processing the images, that is, $11$\% of the power needed to fly. 

To acquire a dataset for a wide-spectrum of flight parameters, we set the drone on a semi-random flight pattern around the edge servers. The pattern is defined by assigning uniformly distributed GPS way-points to the drone in a cylinder of radius equal to $30$m centered on the edge server constellation and confining the altitude in the $[5,15]m$ range. The maximum speed is randomly chosen for every new GPS waypoint between $[1,4]m/s$. A new waypoint is set as soon as the drone reaches $3$~meters from the current one, to obtain a smooth flight as similar as possible to a real application. In drone applications, the outcome of the object detection analysis is promptly needed to take control action and adjust the trajectory. While the action taken after the image analysis is beyond the scope of the current manuscript, we mention target tracking \cite{mueller2016benchmark}, object avoidance \cite{wang2020uav} as possible applications. 

\subsection{Prediction Performance}
\label{sec:exp_results_myopic}

All results are based on an experimental dataset \cite{seremasgithub} collected using the randomized flight patterns described in Section~\ref{sec:exp_setting}.  We first evaluate the prediction performance of the myopic predictors $p^{W,y}_{i+1,n}=\sigma(s_i)$ and associated binary classifier. In other words,  the predictor determines whether at least half of the delay in the future window is below a given threshold, which we set to $\delta^*{=}175$ms. We use the Area Under the Curve (AUC), integral of the ROC with respect to false positives, as performance metric, commonly used to evaluate algorithms predicting an imbalanced target. 

\begin{figure}[h!]
    \centering 
    \includegraphics[width=0.95\columnwidth]{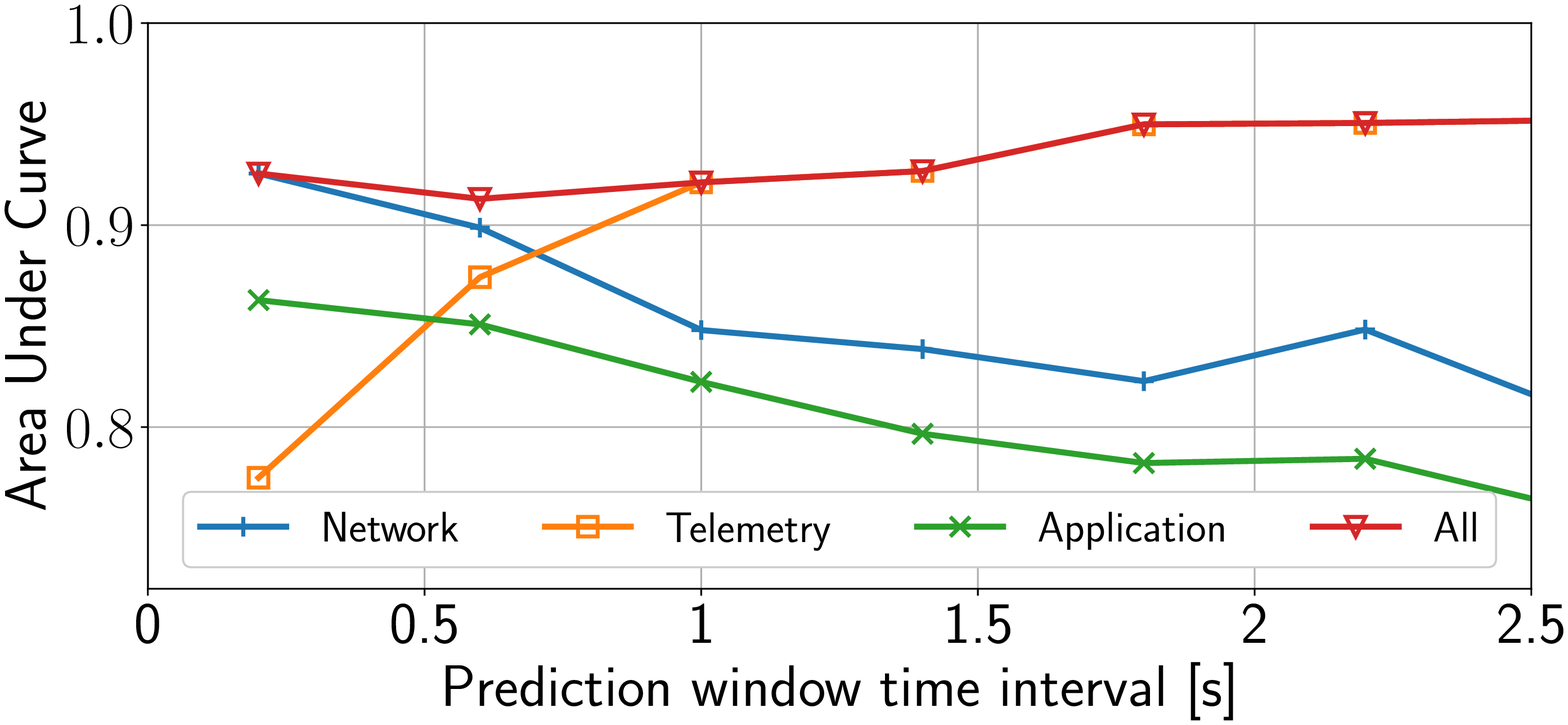}
    \vspace{-2.5mm}
    \caption{Performance of future delay classification for different sets of features. Length of the prediction window is expressed in seconds.}
    \label{fig:future_diff_feats}
    \vspace{-2mm}
\end{figure}

Fig. \ref{fig:future_diff_feats} shows the performance of the predictor trained on different feature blocks as a function of the window $W$ (where we set $y=W/2$). The results highlight how semantic differences across subsets of features influence their predictive power in the short and long term. When the prediction window is small, most of the predictive power lies in networking features, which capture short-term correlations between high delay events. However, network variables struggle to capture longer-term trends, which are, instead predicted by telemetry variables. Indeed, the latter directly influence the distribution of fine-grain network events.

\begin{figure}[h!]
    \centering
    \includegraphics[width=0.95\columnwidth]{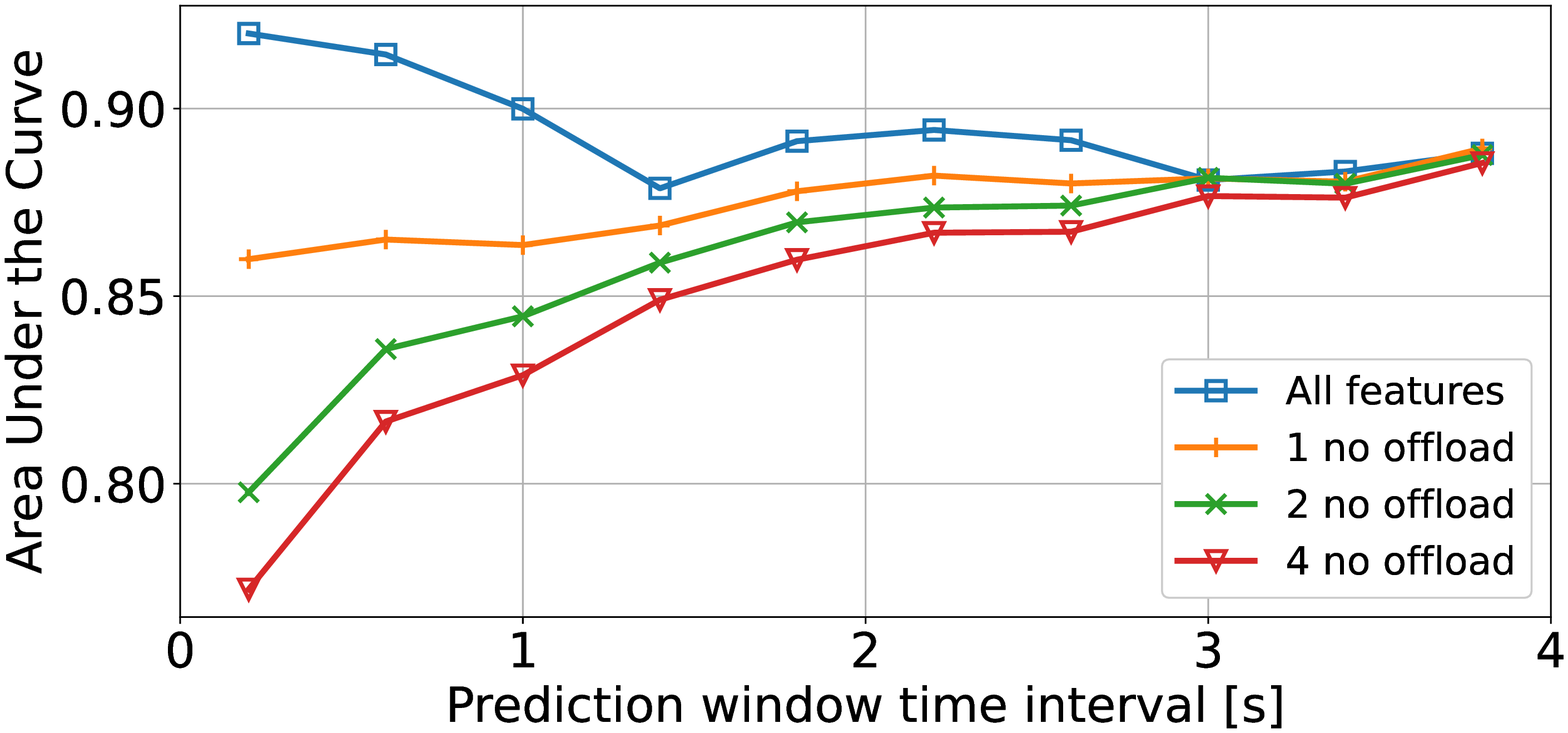}
    \caption{Performance in future delay classification in presence of partial information for recent time slots. Length of the prediction window is expressed in seconds.}
    \label{fig:future_lack_recent}
    \vspace{-2mm}
\end{figure}

As noticed earlier, part of the network information is available only when offloading to a particular edge server. We now analyze how prediction performance is affected when several recent samples lack such information for one server. Fig.~\ref{fig:future_lack_recent} shows how the lack of full state information (which is available only if the edge server is used) in recent samples (last one, last two, etc.) affects the ability of the myopic classifier to accurately predict future pipeline performance as a function of the prediction window $W$ expressed in seconds.
Missing information in one or few recent input samples, has a noticeable effect on classification in the short term, as the AUC reduces by $5\%$ for one sample and $10\%$ for just two samples. On the other hand, as expected, the influence of recent samples fades out when predicting further points in the future. As the decisions of the DRL agent embed the future performance beyond the next delay sample, they also consider the availability of information in future decision instances.

\subsection{Redundant Offloading}
\label{sec:exp_results_redundant}

\begin{figure}[h!]
    \centering
    \includegraphics[width=\columnwidth]{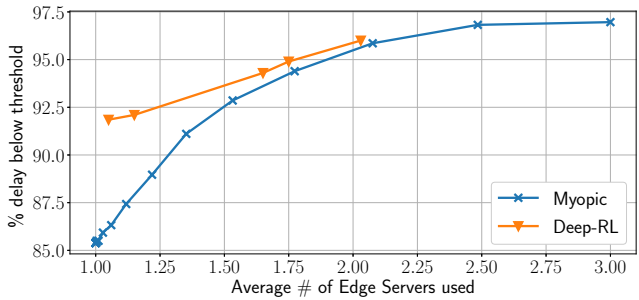}
    \caption{Delay performance and resource utilization trend of the myopic and DRL-based selector.}
\label{fig:channel_latency}
\vspace{-.5em}
\end{figure}

Fig.~\ref{fig:channel_latency} shows the performance of the myopic and DRL selectors in terms of delay (percentage below threshold) and resource usage (average number of edge servers used). The different points for the myopic approach are obtained by varying the parameter $\Delta$, i.e., the bound on the probability that the delay is below threshold. 

The DRL approach, as described in Section \ref{sec:problem_formulation}, generates different points in the plot for different values of the weight $\lambda$ in the cost function, where a larger $\lambda$ favors low delay over resource usage. For comparison, we include a selector which uses all the available edge servers for all the tasks, and a selector which uses the edge server with the best channel quality index. When using all the three edge servers all the time, the myopic selector achieves maximum performance ($\sim 97\%$), whereas when using only one edge server, it achieves $\sim 85.5\%$. We note that a selector that chooses the edge server with the best channel quality achieves $75$\% of tasks with delays below threshold, w.r.t. which we improve $17\%$. Thus, \emph{predictive control greatly improves performance compared to traditional options, even when idealized to task-level granularity without connection delay}.
As we make the bound on $\Delta$ more tight, the myopic approach uses more and more resources.  

We observe that using two edge servers, the myopic controller already achieves a performance roughly $2$\% worse than the three edge server option, demonstrating that prediction can reduce resource usage. However, when using a small amount of resources, the myopic controller's effectiveness sharply decreases. Conversely, the DRL is capable of effectively select small sets of computing pipelines while preserving delay performance. Using $1.1$ edge servers on average, the DRL approach achieves $\sim 92$\%, that is, $7$\% more than the myopic approach. We explain this trend by observing that the DRL agent optimizes the information available to make future decisions, thus maximizing the overall prediction accuracy when resources are scarce and selection needs to be precise.

\begin{figure}[!h]
     \centering
     \begin{subfigure}[b]{0.95\columnwidth}
         \centering
         \includegraphics[trim = 10px 10px 10px 20px, width=\textwidth]{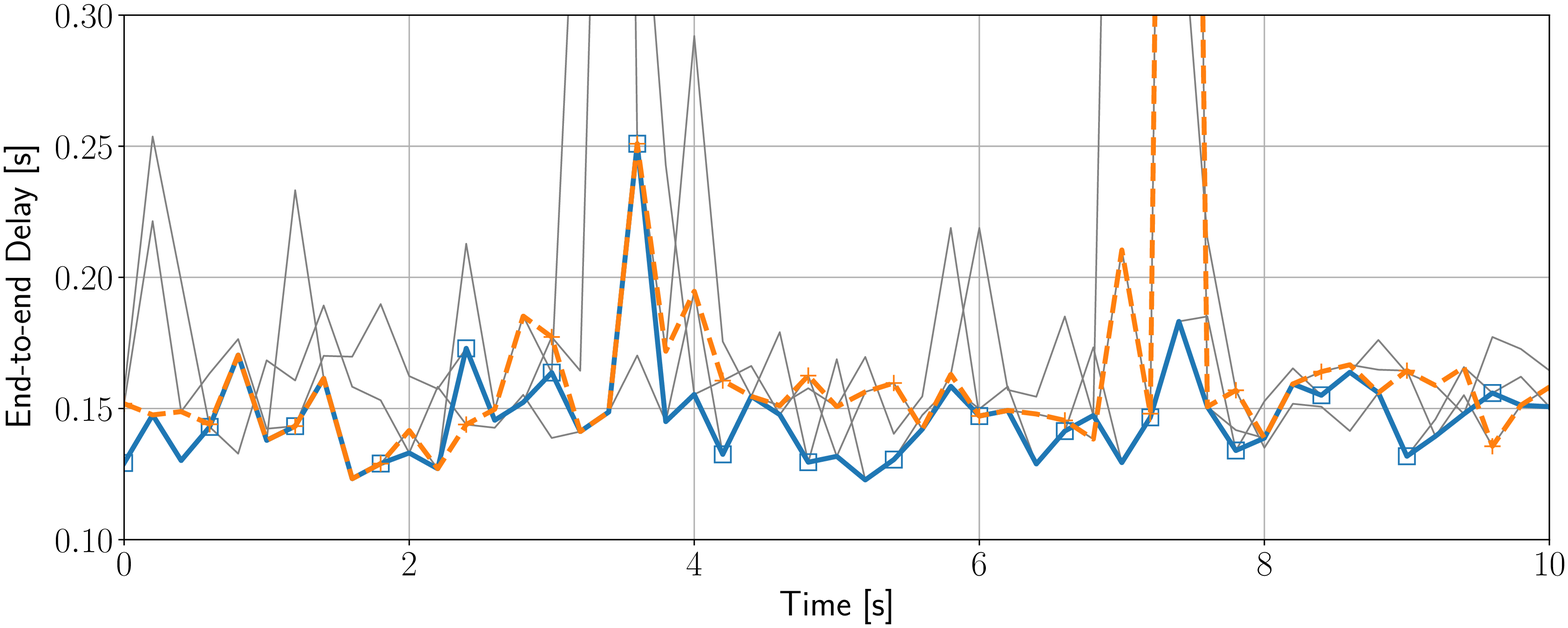}
     \end{subfigure}

     \begin{subfigure}[b]{0.95\columnwidth}
         \centering
         \includegraphics[trim = 10px 10px 10px 20px, width=\textwidth]{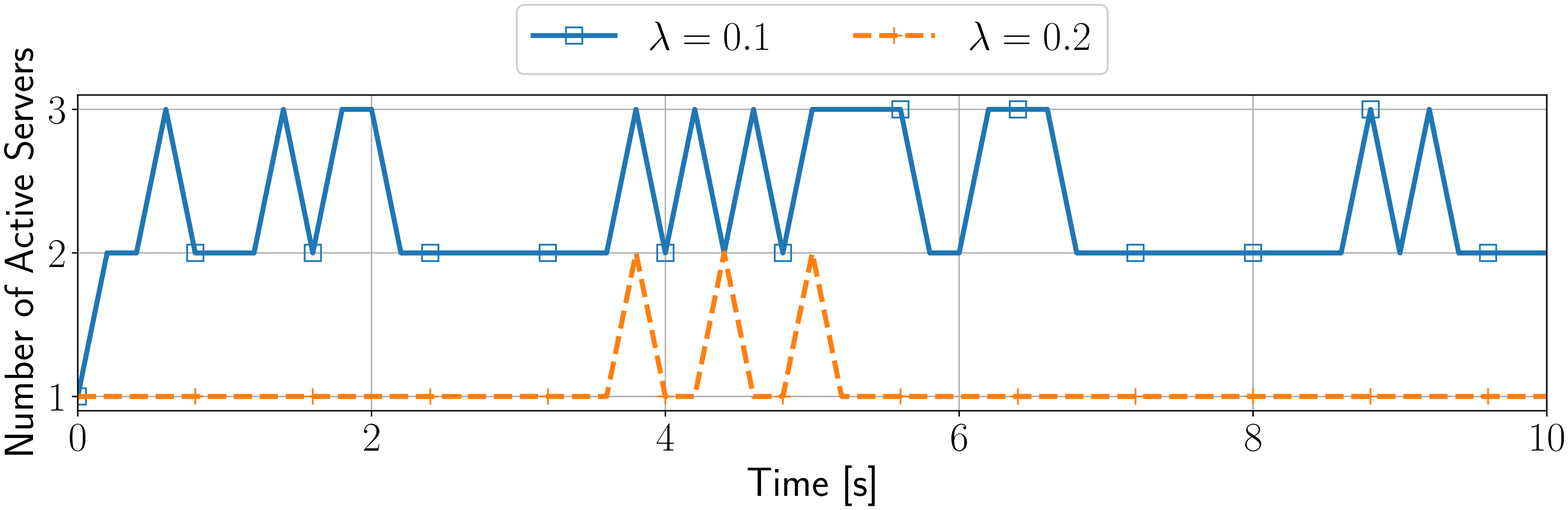}
     \end{subfigure}

     \begin{subfigure}[b]{0.95\columnwidth}
         \centering
         \includegraphics[trim = 10px 10px 10px 20px, width=\textwidth]{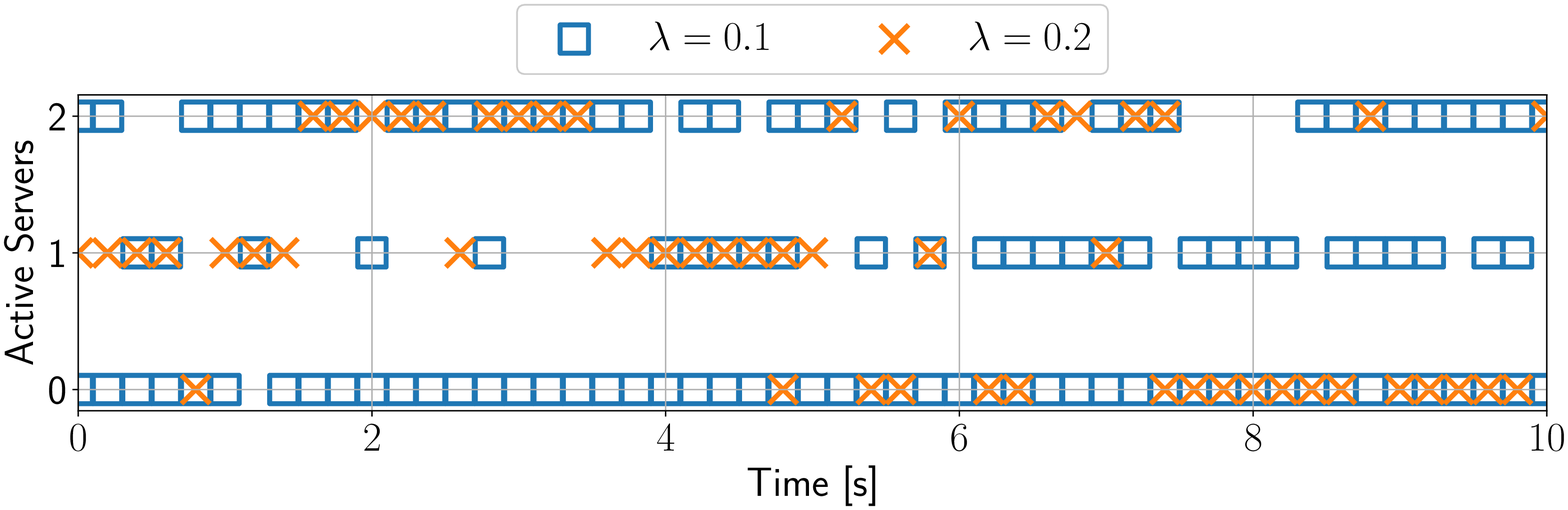}
     \end{subfigure}
    \caption{DRL agent improving delay by using task replication.  We plot in grey the traces of the non-selected delays.\vspace{-0.3cm}}
    \label{fig:rl_trace}
\end{figure}

To further illustrate the behavior of the proposed approach, we show in Figure \ref{fig:rl_trace} a time series of delays and decisions (selected edge servers) of the DRL-based approach for two different $\lambda$ ($0.1$ and $0.2$) used in Eq. \ref{eq:cost}.
We can see that the DRL agent can stabilize delay, where a larger use of resources leads to the avoidance of more delay peaks. We note how the DRL agent rotates the edge servers periodically to harvest information for more informed future decisions.


\section{Related Work}\label{sec:rw}
Edge computing can significantly improve reliability and performance in mobile applications \cite{bonomi2012fog}. Different frameworks perform a multi-layer optimization to exploit the full potential of edge computing \cite{zhou2018computation,cheng2019space}. To fully exploit the edge servers, the user equipment needs to periodically make a decision on whether to process tasks locally, or to offload. In the latter case, there might be multiple technologies or networks available, e.g. \cite{callegaro-baidya-2020-icc}, and a link must be chosen for each transmission. Convex optimization has been proven to be ineffective due to the presence of complex factors such as user's mobility \cite{cao2019intelligent}. Classic approaches  are shown to perform better with coarser granularity settings and when considerable prior knowledge is available. For example, in \cite{9109630} the authors develop an online multi-decision making scheme, solving a task offloading problem while jointly optimizing caching, communication and computation resources in the Internet of Vehicles, exploiting the proximity of users to roadside units. 

Fast-changing mobile networks usually employ data-driven approaches, using Markov Decision Processes (MDP), Q-Learning or DRL. MDPs  achieve a good tradeoff between the flexibility of learning and the data efficiency of a model-based solution  \cite{callegaro-levorato-2021-tvt,zhang2020task}. However,  MDP-based solutions often lead to exceedingly large state spaces, and require vast amounts of data to find the correct transitions for each state-action pair during training. Finally, they are very memory intensive both in training and execution time. For these reasons, DRL approaches have been proposed. Cao et. al \cite{cao2019intelligent} present a general framework for intelligent offloading in multi-access edge computing composed by observation tier, analysis tier, prediction tier and policy tier. In this paper, we consider a much more complicated problem where the trade-off is beyond power efficiency and link performance. Recently, researchers have worked towards simulation environments for drones, for example, OpenUAV \cite{schmittle2018openuav} and FlyNetSim \cite{baidya2018flynetsim}. However, neither of the two environments can capture the interactions between mobility and application delay that are key in this paper. Thus, we are sharing our dataset with the community to further allow research that can explain and exploit these interactions.

\section{Conclusions}
This paper has proposed SeReMAS, a data-driven optimization framework for predictive task offloading in edge-assisted Mobile Autonomous Systems (MASs). We have formulated a Redundant Task Offloading Problem (RTOP) and created a predictor based on Deep Reinforcement Learning (DRL), which produces the optimum task assignment based on application-, network- and telemetry-based features. We have prototyped SeReMAS on a real-world testbed, and extensively evaluated SeReMAS by considering an application where one drone offloads high-resolution images for real-time analysis to three edge servers on the ground. Experimental results show that SeReMAS improves the task execution probability by $17\%$ with respect to existing reactive-based approaches.


\footnotesize
\bibliographystyle{IEEEtran}
\bibliography{reference}
\end{document}